\documentclass[a4paper,11pt,dvipdfmx]{article}
\usepackage{amsfonts}
\usepackage{amssymb}
\usepackage{amsmath}
\usepackage{amsthm}
\usepackage{bm}
\usepackage{here}
\usepackage{tikz}
\usetikzlibrary{intersections,calc,arrows}
\usepackage{color}
\usepackage{cases}
\numberwithin{equation}{section}

\theoremstyle{definition}

\newtheorem{rem}{Remark}[section]
\theoremstyle{plain}

\newtheorem{thm}{Theorem}[section]

\title{Polynomial Hamiltonians for quantum Garnier systems in two variables}

\author{
	Yuichi Ueno
	\footnote{Department of Mathematics, Graduate School of Science, Kobe University, Rokko, Kobe 657-8501, Japan.}
	\footnote{Department of Education, Kogakkan University, Kodakushimoto-cho, Ise 516-8555, Japan.}
	\footnote{E-mail: y-ueno@kogakkan-u.ac.jp}
	}
\date{}

\allowdisplaybreaks
\begin{document}
\maketitle
\renewcommand{\labelenumi}{\rm (\arabic{enumi})}
\begin{abstract} 
We construct and characterize quantum Garnier systems in two variables including degenerate cases by certain holomorphic properties under the quantum canonical transformations. 
\end{abstract}

\section{Introduction}
%
The original Garnier system \cite{G} in $N$ variables is a Hamiltonian system with $N$ time variables obtained from monodromy preserving deformations of a second order Fuchsian ODE on a Riemann sphere $\mathbb{P}^1$ with $N+3$ singular points and $N$ apparent singularities.
The case of $N=1$ coincides with Painlev\'e $\rm VI$ equation.

Degenerate Garnier systems in two variables were constructed by H. Kimura, and they have the following degenerate scheme corresponding to a division of "$5$" \cite{K1,K2,KO}.\\

\begin{picture}(200,70)(0,110)
\put(20,130){\small{$(1,1,1,1,1)$}}
\put(120,130){\small{$(1,1,1,2)$}}
\put(210,130){\small$(1,1,3)$}\put(210,170){\small$(1,2,2)$}
\put(290,130){\small$(1,4)$}\put(290,170){\small$(2,3)$}
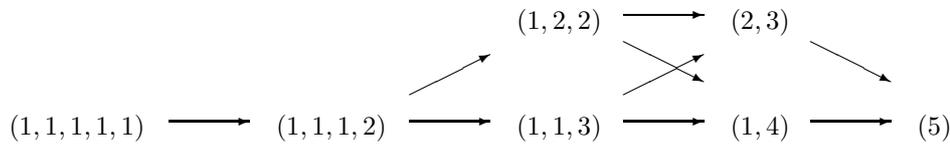
\put(360,130){\small$(5)$}
\put(80,135){\vector(1,0){30}}
\put(170,135){\vector(1,0){30}}\put(170,145){\vector(2,1){30}}
\put(250,135){\vector(1,0){30}}\put(250,175){\vector(1,0){30}}
\put(250,165){\vector(2,-1){30}}\put(250,145){\vector(2,1){30}}
\put(320,135){\vector(1,0){30}}\put(320,165){\vector(2,-1){30}}
\end{picture}

In the works of Takano and his collaborators, the classical Painlev\'e equations were characterized by a geometric method \cite{MMT,M,ST,T}.
The classical Painlev\'e equations can be written as Hamiltonian systems with polynomial Hamiltonian in canonical variables.
Under certain birational canonical transformations, the Hamiltonian systems are transformed again into holomorphic Hamiltonian systems.
Furthermore, it is shown that the classical Painlev\'e  equations can be uniquely characterized by these holomorphic properties.
This is called the "Takano's theory".

In the previous paper \cite{U}, Takano's theory was applied to the quantum Painlev\'e equations.
In the present paper, we extend the results of \cite{U} to quantum Garnier systems.
Namely, we set up suitable quantum canonical transformations and derive quantum Garnier systems in two variables including the degenerate cases uniquely by holomorphic property.
We further show that the flows for two time variables $t_{i} (i=1, 2)$ obtained in this way are commutative.

The organization of this paper is as follows.
In section $2$, we recall the work by Sasano which gives the characterization of the Hamiltonians for the classical Garnier systems by a certain holomorphic property.
In section $3$, we  introduce quantum canonical transformations for Garnier systems in two variables and show that the holomorphic property under the transformations determines the Hamiltonians uniquely (Theorem $3.1.$).
In section $4$, we describe the Hamiltonians determined by the transformations described in section $3$.
Finally, we summarize the results and discuss future works in section 5. 

\section{Garnier system}
Suzuki constructed spaces of initial conditions of the classical Garnier system and its degenerate systems in two variables (namely G(1,1,1,1,1), G(1,1,1,2), G(1,1,3), G(1,2,2), G(1,4), G(2,3) and G(5)) and describe them as symplectic manifolds  \cite{SM1,SM2}.
These systems can be expressed as polynomial Hamiltonian systems on all affine charts.
Based on this fact, a characterization of the Hamiltonians by holomorphic property was considered by Sasano \cite{S2,S3} for the classical Garnier systems  G(1,1,1,1,1), G(1,1,1,2), G(1,1,3), G(1,2,2) and G(1,4).
In this section, we recall Sasano's construction in the case of G(1,1,1,1,1).

The polynomial Hamiltonians of Garnier system G(1,1,1,1,1) were introduced by H. Kimura and K. Okamoto \cite{KO}.
Following the notation by Sasano \cite{S2,S3} (see also \cite{T1,T2,T3,T4}),
we consider the Hamiltonian system of the form
\begin{align} \label{cg}
&dq_{1}=\dfrac{\partial H_{1}}{\partial p_{1}}dt_{1}+\dfrac{\partial H_{2}}{\partial p_{1}}dt_{2},\quad 
dp_{1}=-\dfrac{\partial H_{1}}{\partial q_{1}}dt_{1}-\dfrac{\partial H_{2}}{\partial q_{1}}dt_{2},\notag\\
&dq_{2}=\dfrac{\partial H_{1}}{\partial p_{2}}dt_{1}+\dfrac{\partial H_{2}}{\partial p_{2}}dt_{2},\quad
dp_{1}=-\dfrac{\partial H_{1}}{\partial q_{2}}dt_{1}-\dfrac{\partial H_{2}}{\partial q_{2}}dt_{2},
\end{align}
where the Hamiltonians  $H_{i} (i=1, 2)$  are given by
\begin{equation} \label{cgham}
\begin{array}{ll} 
H_{1}=H_{\rm VI}(q_{1},p_{1}, t ; \alpha_{4}+\alpha_{6},\alpha_{2},\alpha_{1},\alpha_{5},\alpha_{3}) \\
\phantom{H_{1}=}+(2\alpha_{1}+\alpha_{2})\dfrac{q_{1}q_{2}p_{2}}{t_{1}(t_{1}-1)}+\alpha_{3} \{\dfrac{p_{1}}{t_{1}-t_{2}}-\dfrac{(t_{2}-1)p_{2}}{(t_{1}-t_{2})(t_{1}-1)}\}q_{2}+\alpha_{4} \dfrac{t_{2}(p_{2}-p_{1})q_{1}}{t_{1}(t_{1}-t_{2})}\\
\phantom{H_{1}=}+\{\dfrac{2(t_{2}-1)p_{1}p_{2}}{(t_{1}-t_{2})(t_{1}-1)}-\dfrac{t_{1}p_{1}^2+t_{2}p_{2}^2}{t_{1}(t_{1}-t_{2})}+\dfrac{(2q_{1}p_{1}+q_{2}p_{2})p_{2}}{t_{1}(t_{1}-1)}\}q_{1}q_{2}, \\
H_{2}=\pi{(H_{1})},
\end{array}
\end{equation}
and the transformation $\pi$ and the function $H_{\rm VI}$ is given by\footnote{$H_{\rm VI}$ in eq.\eqref{p6ham} is the Hamiltonian for Painlev\'e {\rm VI} if the Fuchs relation
$\tilde\alpha_{0}+\tilde\alpha_{1}+2 \tilde\alpha_{2}+\tilde\alpha_{3}+\tilde\alpha_{4}=1$ is imposed.}
\begin{equation}
\pi : (q_{1}, p_{1}, q_{2}, p_{2}, t_{1}, t_{2} ; \alpha_{1}, \alpha_{2}, \alpha_{3}, \alpha_{4}, \alpha_{5}, \alpha_{6}) \rightarrow (q_{2}, p_{2}, q_{1}, p_{1}, t_{2}, t_{1} ; \alpha_{1}, \alpha_{2}, \alpha_{4}, \alpha_{3}, \alpha_{5}, \alpha_{6}),
\end{equation}
\begin{equation}
\begin{array}{l}  \label{p6ham}
H_{\rm VI}(q, p, t ; \tilde\alpha_{0}, \tilde\alpha_{1}, \tilde\alpha_{2}, \tilde\alpha_{3}, \tilde\alpha_{4}) 
=\dfrac{1}{t(t-1)}\{q(q-1)(q-t)p^2\\[4pt]
\quad 
-\{(\tilde\alpha_{0}-1)q(q-1)+\tilde\alpha_{3}q(q-t)+\tilde\alpha_{4}(q-1)(q-t)\}p+\tilde\alpha_{2}(\tilde\alpha_{1}+\tilde\alpha_{2})(q-t)\}.
\end{array}
\end{equation}

For the Garnier system G(1,1,1,1,1), the canonical transformations considered by Sasano are given as follows: 
\begin{align}\label{tr11111}
r_1:\quad 
&q_{1}=\dfrac{1}{x_1},\quad
p_{1}=-x_{1}^2y_{1}-x_{1}x_{2}y_{2}-\alpha_{1}x_{1},\quad
q_{2}=\dfrac{x_2}{x_1},\quad
p_{2}=x_{1}y_{2},\notag\\
&x_{1}=\dfrac{1}{q_1},\quad
y_{1}=-q_{1}^2p_{1}-q_{1}q_{2}p_{2}-\alpha_{1}q_{1},\quad
x_{2}=\dfrac{q_2}{q_1},\quad
y_{2}=q_{1}p_{2}.\notag\\
r_2:\quad 
&q_{1}=\dfrac{1}{x_1},\quad
p_{1}=-x_{1}^2y_{1}-x_{1}x_{2}y_{2}-\alpha_{2}x_{1},\quad
q_{2}=\dfrac{x_2}{x_1},\quad
p_{2}=x_{1}y_{2},\notag\\
&x_{1}=\dfrac{1}{q_1},\quad
y_{1}=-q_{1}^2p_{1}-q_{1}q_{2}p_{2}-\alpha_{2}q_{1},\quad
x_{2}=\dfrac{q_2}{q_1},\quad
y_{2}=q_{1}p_{2}.\notag\\
r_3:\quad
&q_{1}=-x_{1}y_{1}^2+\alpha_{3}y_{1},\quad
p_{1}=\dfrac{1}{y_{1}},\quad
q_{2}=x_{2},\quad
p_{2}=y_{2},\notag\\
&x_{1}=-q_{1}p_{1}^2+\alpha_{3}p_{1},\quad
y_{1}=\dfrac{1}{p_{1}},\quad
x_{2}=q_{2},\quad
y_{2}=p_{2}.\notag\\
r_4:\quad
&q_{1}=x_{1},\quad
p_{1}=y_{1},\quad
q_{2}=-x_{2}y_{2}^2+\alpha_{4}y_{2},\quad
p_{2}=\dfrac{1}{y_{2}},\notag\\
&x_{1}=q_{1},\quad
y_{1}=p_{1},\quad
x_{2}=-q_{2}p_{2}^2+\alpha_{4}p_{2},\quad
y_{2}=\dfrac{1}{p_{2}}.\notag\\
r_5:\quad
&q_{1}=-x_{1}y_{1}^2-x_{2}+\alpha_{5}y_{1}+1,\quad
p_{1}=\dfrac{1}{y_{1}},\quad
q_{2}=x_{2},\quad
p_{2}=\dfrac{1}{y_{1}}+y_{2},\notag\\
&x_{1}=-q_{1}p_{1}^2-q_{2}p_{1}^2+p_{1}^2+\alpha_{5}p_{1},\quad
y_{1}=\dfrac{1}{p_{1}},\quad
x_{2}=q_{2},\quad
y_{2}=p_{2}-p_{1}.\notag\\
r_6:\quad
&q_{1}=-x_{1}y_{1}^2-\dfrac{t_{1}}{t_{2}}x_{2}+\alpha_{6}y_{1}+t_{1},\quad
p_{1}=\dfrac{1}{y_{1}},\quad
q_{2}=x_{2},\quad
p_{2}=\dfrac{t_{1}}{t_{2}}\dfrac{1}{y_{1}}+y_{2},\notag\\
&x_{1}=-q_{1}p_{1}^2-\dfrac{t_{1}}{t_{2}}q_{2}p_{1}^2+t_{1}p_{1}^2+\alpha_{6}p_{1},\quad
y_{1}=\dfrac{1}{p_{1}},\quad
x_{2}=q_{2},\quad
y_{2}=p_{2}-\dfrac{t_{1}}{t_{2}}p_{1}.
\end{align}

Then  the following was proved by Sasano \cite{S2}.
\begin{thm}\cite{S2}
Consider a polynomial Hamiltonian system with general Hamiltonians $H_{i} (i=1, 2)$ in canonical variables $q_{1}, p_{1}, q_{2}, p_{2}$, and assume the following.
\begin{enumerate}
\item The total degree of the Hamiltonians $H_{i}$ are $5$ in $q_{1}, p_{1}, q_{2}, p_{2}$.
\item Under each transformations $r_{i}(i=1, \ldots, 6)$ of \eqref{tr11111}, the system \eqref{cg} is transformed  into  again a Hamiltonian system with polynomial Hamiltonians.
\end{enumerate}
Then such a system coincides with the system \eqref{cg}-\eqref{p6ham}.
\end{thm}

\begin{rem}
The same is true for the Garnier system in three variables \cite{S3}.
A similar fact is conjectured by Sasano for the Garnier systems in general $n$ variables.
\end{rem}

\section{Quantum Garnier systems in two variables and canonical transformations}
In the following, we consider quantum versions of Garnier systems in two variables.
To properly define them, we consider the following quantum Hamiltonian system of the form

\begin{equation}\label{qgs}
\begin{array}{ll} 
dq_{1}=[H_{1}, p_{1}]dt_{1}+[H_{2}, p_{1}]dt_{2},\hspace{3mm}& dp_{1}=-[H_{1}, q_{1}]dt_{1}-[H_{2}, q_{1}]dt_{2},\\[4pt]
dq_{2}=[H_{1}, p_{2}]dt_{1}+[H_{2}, p_{2}]dt_{2},\hspace{3mm}& dp_{2}=-[H_{1}, q_{2}]dt_{1}-[H_{2}, q_{2}]dt_{2},\\[4pt]
\end{array}
\end{equation}
where $[ , ]$ is the commutator : $[A, B]=AB-BA$ and $q_{1}, p_{1}, q_{2}, p_{2}$ are canonical variables satisfying $[q_{i}, p_{j}]=\delta_{i,j}h\hspace{2mm} (h\in \mathbb{C})$ and $t_{i} (i=1, 2)$ are independent variables of two time evolution.
We will determine the Hamiltonians $H_{i} (i=1, 2)$ by using the holomorphy property.

In order to do this, we need to define quantum canonical transformations in suitable way.
We use the quantum transformations obtained by direct quantization from the Sasano's classical ones for G(1,1,1,1,1), G(1,1,1,2), G(1,1,3), G(1,2,2) and G(1,4) and Suzuki's for G(2,3), G(5).
Though the problems of ambiguity of the ordering of operators arise here, we specify it so that the variables $q_{i}$ are to the left of the variables $p_{j}$.
We note that this specification of ambiguity does not lose generality since the effect of a simple exchange of order can be absorbed in the redefinition of parameters. 
Thus, we will start from the following quantum canonical transformations.
%
\begin{enumerate}
\item 
The case of \rm G(1,1,1,1,1)\\
The same as \eqref{tr11111}.
\item 
The case of \rm G(1,1,1,2)
\begin{align}
r_{1}:\quad
&q_{1}=\dfrac{1}{x_{1}},\quad
p_{1}=-x_{1}^2y_{1}-x_{1}x_{2}y_{2}-\alpha_{3}x_{1},\quad
q_{2}=\dfrac{x_{2}}{x_{1}},\quad
p_{2}=x_{1}y_{2},\notag\\
&x_{1}=\dfrac{1}{q_{1}},\quad
y_{1}=-q_{1}^2p_{1}-q_{1}q_{2}p_{2}-\alpha_{3}q_{1},\quad
x_{2}=\dfrac{q_{2}}{q_{1}},\quad
y_{2}=q_{1}p_{2},\notag\\
r_{2}:\quad
&q_{1}=\dfrac{1}{x_1},\quad
p_{1}=-x_{1}^2y_{1}-x_{1}x_{2}y_{2}-\alpha_{4}x_{1}, \quad
q_{2}=\dfrac{x_2}{x_1},\quad
p_{2}=x_{1}y_{2},\notag\\
&x_{1}=\dfrac{1}{q_1},\quad
y_{1}=-q_{1}^2p_{1}-q_{1}q_{2}p_{2}-\alpha_{4}q_{1}, \quad
x_{2}=\dfrac{q_2}{q_1},\quad
y_{2}=q_{1}p_{2},\notag\\
r_{3}:\quad
&q_{1}=x_{1},\quad
p_{1}=\eta (\dfrac{1}{x_{1}})^2(x_{2}-1)-\alpha_{1}\dfrac{1}{x_{1}}+y_{1}, \quad
q_{2}=x_{2},\quad
p_{2}=-\eta \dfrac{1}{x_{1}}+y_{2},\notag\\
&x_{1}=q_{1},\quad
y_{1}=\eta (\dfrac{1}{q_{1}})^2(q_{2}+1)+\alpha_{1}\dfrac{1}{q_{1}}+p_{1},\quad
x_{2}=q_{2},\quad
y_{2}=\eta \dfrac{1}{q_{1}}+p_{2},\notag\\
r_{4}:\quad
&q_{1}=x_{1},\quad
p_{1}=y_{1},\quad
q_{2}=-x_{2}y_{2}^2+\alpha_{2}y_{2},\quad
p_{2}=\dfrac{1}{y_{2}},\notag\\
&x_{1}=q_{1},\quad
y_{1}=p_{1},\quad
x_{2}=-q_{2}p_{2}^2+\alpha_{2}p_{2},\quad
y_{2}=\dfrac{1}{p_{2}},\notag\\
r_{5}:\quad
&q_{1}=-x_{1}y_{1}^2-\dfrac{t_{1}}{t_{2}}x_{2}+\alpha_{5}y_{1}+t_{1},\quad
p_{1}=\dfrac{1}{y_{1}},\quad
q_{2}=x_{2},\quad
p_{2}=\dfrac{t_{1}}{t_{2}}\dfrac{1}{y_{1}}+y_{2},\notag\\
&x_{1}=-q_{1}p_{1}^2-\dfrac{t_{1}}{t_{2}}q_{2}p_{1}^2+t_{1}p_{1}^2+\alpha_{5}p_{1},\quad
y_{1}=\dfrac{1}{p_{1}},\quad
x_{2}=q_{2},\quad
y_{2}=-\dfrac{t_{1}}{t_{2}}{p_{2}}+p_{2}.\label{tr1112} 
\end{align}
\item 
The case of \rm G(1,1,3)
\begin{align}
r_{1}:\quad
&q_{1}=-x_{1}y_{1}^2-x_{2}y_{1}y_{2}+\alpha_{1}y_{1},\quad
p_{1}=\dfrac{1}{y_{1}}, \quad
q_{2}=x_{2}y_{1},\quad
p_{2}=\dfrac{y_{2}}{y_{1}},\notag\\
&x_{1}=-q_{1}p_{1}^2-q_{2}p_{1}p_{2}+\alpha_{1}p_{1},\quad
y_{1}=\dfrac{1}{p_{1}}, \quad
x_{2}=q_{2}p_{1},\quad
y_{2}=\dfrac{p_{2}}{p_{1}},\notag\\
r_{2}:\quad
&q_{1}=\dfrac{1}{x_{1}},\quad
p_{1}=-x_{1}^2y_{1}-\alpha_{2}x_{1}, \quad
q_{2}=x_{2},\quad
p_{2}=y_{2},\notag\\
&x_{1}=\dfrac{1}{q_{1}},\quad
y_{1}=-q_{1}^2p_{1}-\alpha_{2}q_{1}, \quad
x_{2}=q_{2},\quad
y_{2}=p_{2},\notag\\
r_{3}:\quad
&q_{1}=x_{1},\quad
p_{1}=y_{1}, \quad
q_{2}=\dfrac{1}{x_{2}},\quad
p_{2}=-x_{2}^2y_{2}-\alpha_{3}x_{2},\notag\\
&x_{1}=q_{1},\quad
y_{1}=p_{1}, \quad
x_{2}=\dfrac{1}{q_{2}},\quad
y_{2}=-q_{2}^2p_{2}-\alpha_{3}q_{2},\notag\\
r_{4}:\quad
&q_{1}=-x_{1}y_{1}^2-x_{2}y_{1}y_{2}+\dfrac{2}{y_{1}}(y_{2}+1)+\alpha_{4}y_{1}-2t_{1},\quad
p_{1}=\dfrac{1}{y_{1}}, \notag\\
&q_{2}=x_{2}y_{1}+\dfrac{2}{y_{1}}(y_{2}+1)-2t_{2},\quad
p_{2}=\dfrac{y_{2}}{y_{1}}, \notag\\
&x_{1}=-q_{1}p_{1}^2-q_{2}p_{1}p_{2}+2p_{1}^3+4p_{1}^2p_{2}+2p_{1}p_{2}^2,\quad
y_{1}=2t_{1}p_{1}^2-2t_{2}p_{1}p_{2}+\alpha_{4}p_{1}+\dfrac{1}{p_{1}},\notag\\
&x_{2}=q_{2}p_{1}-2p_{1}^2-2p_{1}p_{2}+2t_{2}p_{1},\quad
y_{2}=\dfrac{p_{2}}{p_{1}}.\label{tr113}
\end{align}
\item 
The case of \rm G(1,2,2)
\begin{align}
r_{1}:\quad
&q_{1}=\dfrac{1}{x_1},\quad
p_{1}=-x_{1}y_{1}^2-\alpha_{1}x_{1}, \quad
q_{2}=x_{2},\quad
p_{2}=y_{2},\notag\\
&x_{1}=\dfrac{1}{q_1},\quad
y_{1}=-q_{1}p_{1}^2-\alpha_{1}q_{1}, \quad
x_{2}=q_{2},\quad
y_{2}=p_{2},\notag\\
r_{2}:\quad
&q_{1}=\dfrac{1}{x_{1}},\quad
p_{1}=-x_{1}^2y_{1}-\alpha_{2}x_{1}-y_{2}+1, \quad
q_{2}=\dfrac{1}{x_{1}}+x_{2},\quad
p_{2}=y_{2},\notag\\
&x_{1}=\dfrac{1}{q_{1}},\quad
y_{1}=-{q_1}^2p_{1}-\alpha_{2}q_{1}-p_{2}+1, \quad
x_{2}=q_{2}-q_{1},\quad
y_{2}=p_{2},\notag\\
r_{3}:\quad
&q_{1}=x_{1},\quad
p_{1}=y_{1}, \quad
q_{2}=\dfrac{1}{x_{2}},\quad
p_{2}=-x_{2}^2y_{2}-\alpha_{3}x_{2},\notag\\
&x_{1}=q_{1},\quad
y_{1}=p_{1}, \quad
x_{2}=\dfrac{1}{q_{2}},\quad
y_{2}=-q_{2}^2p_{2}-\alpha_{3}q_{2},\notag\\
r_{4}:\quad
&q_{1}=x_{1},\quad
p_{1}=-\dfrac{t_{2}}{t_{1}}(\dfrac{1}{x_{1}})^2y_{2}-2\dfrac{x_{2}}{x_{1}}y_{2}-t_{1}(\dfrac{1}{x_{1}})^2+2\alpha_{4}\dfrac{1}{x_{1}}+y_{1}, \notag\\
&q_{2}=x_{1}^2x_{2}+\dfrac{t_{2}}{t_{1}}x_{1},\quad
p_{2}=(\dfrac{1}{x_{1}})^2y_{2}, \notag\\
&x_{1}=q_{1},\quad
y_{1}=-2\dfrac{q_{2}}{q_{1}}p_{2}+t_{1}(\dfrac{1}{q_{1}})^2-2\alpha_{4}\dfrac{1}{q_{1}}+p_{1}-\dfrac{t_{2}}{t_{1}}p_{2}, \notag\\
&x_{2}=(\dfrac{1}{q_{1}})^2q_{2}-\dfrac{t_{2}}{t_{1}}\dfrac{1}{q_{1}},\quad
y_{2}=q_{1}^2p_{2}.\label{tr122}
\end{align}
\item 
The case of \rm G(1,4)
\begin{align}
r_{1}:\quad
&q_{1}=\dfrac{1}{x_1},\quad
p_{1}=-x_{1}^2y_{1}-\alpha_{1}x_{1}, \quad
q_{2}=x_{2},\quad
p_{2}=y_{2},\notag\\
&x_{1}=\dfrac{1}{q_1},\quad
y_{1}=-q_{1}^2p_{1}-\alpha_{1}q_{1}, \quad
x_{2}=q_{2},\quad
y_{2}=p_{2},\notag\\
r_{2}:\quad
&q_{1}=x_{1},\quad
p_{1}=y_{1}, \quad
q_{2}=\dfrac{1}{x_{2}},\quad
p_{2}=-x_{2}^2y_{2}-\alpha_{2}x_{2},\notag\\
&x_{1}=q_{1},\quad
y_{1}=p_{1}, \quad
x_{2}=\dfrac{1}{q_{2}},\quad
y_{2}=-q_{2}^2p_{2}-\alpha_{2}q_{2},\notag\\	
r_{3}:\quad
&q_{1}=\dfrac{1}{x_{1}},\quad
p_{1}=-x_{1}^2y_{1}+2x_{1}x_{2}y_{2}-(\dfrac{1}{x_{1}})^2y_{2}+2(\dfrac{1}{x_{1}})^2 -\alpha_{3}x_{1}-\dfrac{t_{1}-t_{2}}{2}y_{2}+t_{1}, \notag\\
&q_{2}=x_{1}^2x_{2}+\dfrac{t_{1}-t_{2}}{2}x_{1}+\dfrac{1}{x_{1}},\quad
p_{2}=(\dfrac{1}{x_{1}})^2y_{2}, \notag\\
&x_{1}=\dfrac{1}{q_{1}},\quad
y_{1}=2q_{1}^4-q_{1}^2p_{1}-3q_{1}^2p_{2}+2q_{1}q_{2}p_{2}+t_{1}q_{1}^2-\alpha_{3}x_{1}+\dfrac{t_{1}-t_{2}}{2}p_{2}, \notag\\
&x_{2}=-q_{1}^3+q_{1}^2q_{2}-\dfrac{t_{1}-t_{2}}{2}q_{1},\quad
y_{2}=(\dfrac{1}{q_{1}})^2p_{2}.\label{tr14}
\end{align}
\item 
The case of \rm G(2,3)
\begin{align}
r_{1}:\quad
&q_{1}=\dfrac{1}{x_1},\quad
p_{1}=-x_{1}^2y_{1}-x_{1}x_{2}y_{2}-\alpha_{1}x_{1}, \quad
q_{2}=\dfrac{x_{2}}{x_{1}},\quad
p_{2}=x_{1}y_{2},\notag\\
&x_{1}=\dfrac{1}{q_1},\quad
y_{1}=-q_{1}^2p_{1}-q_{1}q_{2}p_{2}-\alpha_{1}q_{1}, \quad
x_{2}=\dfrac{q_{2}}{q_{1}},\quad
y_{2}=q_{1}p_{2},\notag\\
r_{2}:\quad
&q_{1}=\dfrac{x_{1}}{x_{2}},\quad
p_{1}=x_{2}y_{1}, \quad
q_{2}=\dfrac{1}{x_{2}},\quad
p_{2}=-x_{2}^2y_{2}-x_{1}x_{2}y_{1}-\alpha_{1}x_{2},\notag\\
&x_{1}=\dfrac{q_{1}}{q_{2}},\quad
y_{1}=q_{2}p_{1}, \quad
x_{2}=\dfrac{1}{q_{2}},\quad
y_{2}=-q_{2}^2p_{2}-q_{1}q_{2}p_{1}-\alpha_{1}q_{2},\notag\\	
r_{3}:\quad
&q_{1}=x_{1},\quad
p_{1}=-\eta t_{1} \dfrac{1}{x_{2}}+y_{1}, \quad
q_{2}=x_{2},\quad
p_{2}=\eta t_{1} (\dfrac{1}{x_{2}})^2-\eta t_{1}t_{2}(\dfrac{1}{x_{2}})^2+\alpha_{2}\dfrac{1}{x_{2}}+y_{2},\notag\\
&x_{1}=q_{1},\quad
y_{1}=\eta t_{1}\dfrac{1}{q_{2}}+p_{1}, \quad
x_{2}=q_{2},\quad
y_{2}=-\eta t_{1}q_{1}(\dfrac{1}{q_{2}})^2+\eta t_{1}t_{2}(\dfrac{1}{q_{2}})^2-\alpha_{2}\dfrac{1}{q_{2}}+p_{2},\notag\\
r_{4}:\quad
&q_{1}=\dfrac{1}{x_{1}},\quad
p_{1}=-x_{1}^2y_{1}-x_{1}x_{2}y_{2}-(\alpha_{1}+\alpha_{2})x_{1}-\eta t_{1}\dfrac{x_{1}}{x_{2}},\notag\\
&q_{2}=\dfrac{x_{2}}{x_{1}},\quad
p_{2}=-\eta t_{1} (\dfrac{x_{1}}{x_{2}})^2+\eta t_{1}x_{1}(\dfrac{1}{x_{2}})^2+x_{1}y_{2}+\alpha_{2}\dfrac{x_{1}}{x_{2}},\notag\\
&x_{1}=\dfrac{1}{q_{1}},\quad
y_{1}=-q_{1}^2p_{1}-q_{1}q_{2}p_{2}-\eta t_{1}t_{2}\dfrac{q_{1}}{q_{2}}-\alpha_{1}q_{1},\notag\\
&x_{2}=q_{2},\quad
y_{2}=\eta t_{1}t_{2}q_{1}(\dfrac{1}{q_{2}})^2-\eta t_{1}(\dfrac{q_{1}}{q_{2}})^2+q_{1}p_{2}-\alpha_{2}\dfrac{q_{1}}{q_{2}},\notag\\
r_{5}:\quad
&q_{1}=\dfrac{1}{x_{1}},\quad
p_{1}=-x_{1}^2y_{1}-x_{1}x_{2}y_{2}-(1+\alpha_{1}-\alpha_{2}+2\alpha_{3})x_{1}+\dfrac{1}{2x_{1}},\notag\\
&q_{2}=\dfrac{x_{2}}{x_{1}},\quad
p_{2}=x_{1}y_{2}-\dfrac{1}{2},\notag\\
&x_{1}=\dfrac{1}{q_{1}},\quad
y_{1}=\dfrac{1}{2}q_{1}^3-q_{1}^2p_{1}-q_{1}q_{2}p_{2}-\dfrac{1}{2}q_{1}q_{2}-(1+\alpha_{1}-\alpha_{2}+2\alpha_{3})q_{1},\notag\\
&x_{2}=\dfrac{q_{1}}{q_{2}},\quad
y_{2}=q_{1}p_{2}-\dfrac{1}{2},\notag\\
r_{6}:\quad
&q_{1}=\dfrac{x_{1}}{x_{2}},\quad
p_{1}=\dfrac{x_{1}}{2x_{2}}+x_{2}y_{1},\notag\\
&q_{2}=\dfrac{1}{x_{2}},\quad
p_{2}=-x_{1}x_{2}^2y_{2}-x_{1}x_{2}y_{1}-(1+\alpha_{1}-\alpha_{2}+2\alpha_{3})x_{2}-\dfrac{1}{2},\notag\\
&x_{1}=\dfrac{q_{1}}{q_{2}},\quad
y_{1}=-\dfrac{1}{2}q_{1}q_{2}+q_{2}p_{1},\notag\\
&x_{2}=\dfrac{1}{q_{2}},\quad
y_{2}=\dfrac{1}{2}q_{1}^2q_{2}-q_{1}q_{2}p_{1}-q_{2}^2p_{2}-\dfrac{1}{2}q_{2}^2-(1+\alpha_{1}-\alpha_{2}+2\alpha_{3})q_{2}.\label{tr23}
\end{align}
\item 
The case of \rm G(5)
\begin{align}
r_{1}:\quad
&q_{1}=\dfrac{1}{x_1},\quad
p_{1}=-x_{1}^2y_{1}-x_{1}x_{2}y_{2}-\alpha_{1}x_{1}, \quad
q_{2}=\dfrac{x_{2}}{x_{1}},\quad
p_{2}=x_{1}y_{2},\notag\\
&x_{1}=\dfrac{1}{q_1},\quad
y_{1}=-q_{1}^2p_{1}-q_{1}q_{2}p_{2}-\alpha_{1}q_{1}, \quad
x_{2}=\frac{q_{2}}{q_{1}},\quad
y_{2}=q_{1}p_{2},\notag\\
r_{2}:\quad
&q_{1}=\frac{x_{1}}{x_{2}},\quad
p_{1}=x_{2}y_{1}, \quad
q_{2}=\dfrac{1}{x_{2}},\quad
p_{2}=-x_{1}x_{2}y_{1}-x_{2}^2y_{2}-\alpha_{1}x_{2},\notag\\
&x_{1}=\frac{q_{1}}{q_{2}},\quad
y_{1}=q_{2}p_{1}, \quad
x_{2}=\dfrac{1}{q_{2}},\quad
y_{2}=-q_{1}q_{2}p_{1}-q_{2}^2p_{2}-\alpha_{1}q_{2},\notag\\	
r_{3}:\quad
&q_{1}=\dfrac{1}{x_{1}},\quad
p_{1}=-2(\dfrac{x_{2}}{x_{1}})^2-x_{1}^2y_{1}-x_{1}x_{2}y_{2}-(\alpha_{1}-2\alpha_{2})x_{1}+2\dfrac{1}{x_{1}}-2t_{2},\notag\\
&q_{2}=\dfrac{x_{2}}{x_{1}},\quad
p_{2}=2(\dfrac{x_{2}}{x_{1}})^3-4(\dfrac{1}{x_{1}})^2x_{2}+x_{1}y_{2}-2t_{1}, \notag\\
&x_{1}=\dfrac{1}{q_{1}},\quad
y_{1}=2q_{1}q_{2}^4-6q_{1}^2q_{2}^2+2q_{1}^3-q_{1}^2p_{1}-2q_{1}q_{2}p_{2}-2t_{2}q_{1}^2-2tq_{1}q_{2}-(\alpha_{1}-2\alpha_{3})q_{1},\notag\\
&x_{2}=\dfrac{q_{2}}{q_{1}},\quad
y_{2}=-2q_{1}q_{2}^3+4q_{1}^2q_{2}+q_{1}p_{2}+2t_{1}q_{1},\notag\\
r_{4}:\quad
&q_{1}=\dfrac{x_{1}}{x_{2}},\quad
p_{1}=2\dfrac{x_{1}}{x_{2}}-2(\dfrac{1}{x_{2}})^2+x_{2}y_{1}-2t_{2},\notag\\
&q_{2}=\dfrac{1}{x_{2}},\quad
p_{2}=-4x_{1}(\dfrac{1}{x_{2}})^2-x_{1}x_{2}y_{1}+2(\dfrac{1}{x_{2}})^3-x_{2}^2y_{2}-(\alpha_{1}-2\alpha_{2})x_{2}-2t_{1},\notag\\
&x_{1}=\dfrac{q_{1}}{q_{2}},\quad
y_{1}=2q_{2}^3-2q_{1}q_{2}+q_{2}p_{1}+2t_{2}q_{2}, \notag\\
&x_{2}=\dfrac{1}{q_{2}},\quad
y_{2}=2q_{2}^5-6q_{1}q_{2}^3+2q_{1}^2q_{2}-q_{1}q_{2}p_{1}-q_{2}^2p_{2}-2t_{2}q_{1}q_{2}-2t_{1}q_{2}^2-(\alpha_{1}-2\alpha_{2})q_{2}.\label{tr5}
\end{align}
\end{enumerate}

\begin{rem}
We note here the relationship between the \cite{S2, S3}, \cite{SM2} and this paper with respect to the parameters and variables.
\begin{enumerate}
\item 
The case of \rm G(1,1,1,1,1)\\
The $(\alpha_{1}+\alpha_{2}, t, s)$ in \cite{S2} correspond to $(\alpha_{2}, t_{1}, t_{2})$ in this paper.
\item 
The case of \rm G(1,1,1,2)\\
The $(\nu, \alpha_{0}, t, s)$ in \cite{S3} correspond to $(\alpha_{4}, \alpha_{5}, t_{1}, t_{2})$ in this paper.
\item
The case of \rm G(1,1,3)\\
The $(t, s)$ in \cite{S3} correspond to $(t_{1}, t_{2})$ in this paper.
\item
The case of \rm G(1,2,2)\\
The $(\alpha_{0}, \alpha_{3}, \alpha_{2}, \alpha_{1},t ,s)$ in \cite{S3} correspond to $(\alpha_{1}, \alpha_{2}, \alpha_{3}, \alpha_{4}, t_{1}, t_{2})$ in this paper.
\item 
The case of \rm G(1,4)\\
The $(\alpha_{3}, \alpha_{1}, \alpha_{2}, t, s)$ in \cite{S3} correspond to $(\alpha_{1}, \alpha_{2}, \alpha_{3}, t_{1}, t_{2})$ in this paper.
\item 
The case of \rm G(2,3)\\
The $(\nu, \alpha_{0}, \alpha_{\infty}, s_{2}, s_{1})$ in \cite{SM2} correspond to $(\alpha_{1}, \alpha_{2}, \alpha_{3}, t_{1}, t_{2})$ in this paper.\\
The variables $(q_{i}^{02}, p_{i}^{02})$ in Theorem 6 of \cite{SM2} correspond to $(x_{i}, y_{i})$ in $r_{4}$ in this paper.\\
The variables $(q_{i}^{\infty 1}, p_{i}^{\infty 1})$ in Theorem 6 of \cite{SM2} correspond to $(x_{i}, y_{i})$ in $r_{5}$ in this paper.\\
The variables $(q_{i}^{\infty 2}, p_{i}^{\infty 2})$ in Theorem 6 of \cite{SM2} correspond to $(x_{i}, y_{i})$ in $r_{6}$ in this paper.
\item 
The case of \rm G(5)\\
The $(\nu, \alpha, s_{2}, s_{1})$ in \cite{SM2} correspond to $(\alpha_{1}, \alpha_{2}, t_{1}, t_{2})$ in this paper.\\
The variables $(q_{i}^{\infty 1}, p_{i}^{\infty 1})$ in Theorem 7 of \cite{SM2} correspond to $(x_{i}, y_{i})$ in $r_{3}$ in this paper.\\
The variables $(q_{i}^{\infty 2}, p_{i}^{\infty 2})$ in Theorem 7 of \cite{SM2} correspond to $(x_{i}, y_{i})$ in $r_{4}$ in this paper.
\end{enumerate}
\end{rem}

For each case (1)-(7), the following is true.

\begin{thm}\label{thm}
Consider a Hamiltonian system \eqref{qgs} with noncommutative polynomial Hamiltonians $H_{i} (i=1, 2)$ in quantum canonical variables $q_{1}, p_{1}, q_{2}, p_{2}$, and assume the following.
\begin{enumerate}
\item The total degree of the Hamiltonians $H_{i}$ are $5$ in $q_{1}, p_{1}, q_{2}, p_{2}$.
\item Under the corresponding transformations $r_{i}$, the system \eqref{qgs} are transformed  into again a Hamiltonian system with polynomial Hamiltonians.
\end{enumerate}
Then such a system is determined uniquely.
\end{thm}

\noindent {\it Proof.} 
The proof is based on explicit calculation.
In this calculation, we need commutation relations between canonical variables such as $[q,p]=h$ including there inverses.
Which can be computed as follows:
\begin{equation}
[p, q^{-1}]=h q^{-2},\quad [p^{-1}, q]=h p^{-2}.
\end{equation}
Fortunately we do not need the commutator such as $[p^{i}, q^{j}]$ for $i,  j<0$ in our computation.

As an example, we consider the Hamiltonian $H_{1}$ for $t$-flow in case of G(1,1,1,1,1) in \eqref{qgs}.
We put the Hamiltonian as follows:
\begin{equation}
H_{1}=\sum_{i_1, i_2, i_3, i_4}k_{i_1, i_2, i_3, i_4}q_1^{i_1} p_1^{i_2} q_2^{i_3} p_2^{i_4},
\end{equation}
where sum is taken over nonnegative integers such that $i_1+i_2+i_3+i_4 \leq5$.
Since the transformations $r_{1},\ldots,r_{5}$ do not contain  the variable $t_{1}$, the transformed equation can be computed simply by looking at the transformation of the Hamiltonian $H_{1}$.
For example, applying the transformation $r_{1}$
\begin{align}\label{cg1}
r_{1}:\quad
q_{1}=\dfrac{1}{x_1},\quad
p_{1}=-x_{1}^2y_{1}-x_{1}x_{2}y_{2}-\alpha_{1}x_{1},\quad
q_{2}=\dfrac{x_2}{x_1},\quad
p_{2}=x_{1}y_{2},
\end{align}
to $H_{1}$, 
we get a rational expression of $x_{1}, y_{1}, x_{2}, y_{2}$  with poles $x_{1}^{-1}$ to $x_{1}^{-5}$, 
\begin{equation}
r_{1}(H_{1})=(k_{0,2,1,0}-k_{1,1,1,0}+(h-\alpha_{1})k_{1,2,1,1}-2(h-\alpha_{1})k_{2,1,2,0})\dfrac{1}{x_{1}}x_{2}^2y_{2}+\cdots.
\end{equation}
We impose the condition that all such coefficients of the pole terms vanish.
The holomorphy conditions arising from the transformations $r_{2}, \ldots, r_{5}$ are similar.
Solving the holomorphic conditions for $r_{1}, \ldots, r_{5}$, the unknown coefficients $k_{i_1, i_2, i_3, i_4}$ can be determined in terms of five free parameters.
For the transformation $r_6$ which contains the time variable $t_{1}$, one should compute the holomorphic condition by looking at the equation for the $t_{1}$-flow
\begin{equation}
\frac{df}{dt_{1}}=[H_{1}, f]+\frac{\partial f}{\partial t_{1}},
\quad 
f=x_{1}, y_{1}, x_{2}, y_{2}
\end{equation} 
The right hand sides can be written as rational expressions of $x_{1}, y_{1}, x_{2}, y_{2}$ and we require that they are holomorphic.
In this way, the remaining five unknown coefficients can be determined uniquely. 
The same can be shown for the Hamiltonian $H_{2}$.
The degenerate cases are similar. 
For each case, the Hamiltonians obtained in this way are presented in the next section.
\qed

\section{Determined Hamiltonians}\label{DM}
Below we describe the Hamiltonians $H_{i} (i=1, 2)$ determined by the holomorphy under the quantum canonical transformation in \S3.
\begin{enumerate}
\item
The case of \rm G(1,1,1,1,1)\\[4pt]
{\bf The Hamiltonian $H_1$ for $t_{1}$-flow.}
\begin{equation}
\begin{array}{l}
H_{1}=\dfrac{1}{(h-\alpha_{1}-\alpha_{2}-\alpha_{3}-\alpha_{4}-\alpha_{5}-\alpha_{6})t_{1}(t_{1}-1)(t_{1}-t_{2})}\{(t_{2}-t_{1})q_{1}^3p_{1}^2+2(t_{2}-t_{1})q_{1}^2q_{2}p_{1}p_{2}\\[10pt]
\phantom{H_{1}=}+(t_{2}-t_{1})q_{1}q_{2}^2p_{2}^2+(t_{1}+1)(t_{1}-t_{2})q_{1}^2p_{1}^2-2t_{1}(t_{2}-1)q_{1}q_{2}p_{1}p_{2}+t_{1}(t_{1}-1)q_{1}q_{2}p_{1}^2\\[10pt]
\phantom{H_{1}=}+t_{2}(t_{1}-1)q_{1}q_{2}p_{2}^2-(h-\alpha_{1}-\alpha_{2})(t_{2}-t_{1})q_{1}(q_{1}p_{1}+q_{2}p_{2})-t_{1}(t_{1}-t_{2})q_{1}p_{1}^2\\[10pt]
\phantom{H_{1}=}+(h(t_{2}-t_{1})+(\alpha_{1}+\alpha_{2})(t_{1}-t_{2})+\alpha_{3}t_{1}(t_{2}-t_{1})+\alpha_{4}t_{2}(t_{1}+1)\\[10pt]
\phantom{H_{1}=}-\alpha_{5}(t_{1}^2-t_{1}+t_{2}-t_{1}t_{2})q_{1}p_{1}-\alpha_{4}t_{2}(t_{1}-1)q_{1}p_{2}-\alpha_{3}t_{1}(t_{1}-1)q_{2}p_{1}\\[10pt]
\phantom{H_{1}=}+\alpha_{3}t_{1}(t_{2}-1)q_{2}p_{2}+\alpha_{1}\alpha_{2}(t_{2}-t_{1})q_{1}+\alpha_{3}t_{1}(t_{1}-t_{2})p_{1}\}. 
\label{hg11111t1}
\end{array}
\end{equation}

{\bf The Hamiltonian $H_2$ for $t_{2}$-flow.}
\begin{equation}
\begin{array}{l}
H_{2}=\dfrac{1}{(h-\alpha_{1}-\alpha_{2}-\alpha_{3}-\alpha_{4}-\alpha_{5}-\alpha_{6})t_{2}(t_{2}-1)(t_{2}-t_{1})}\{(t_{1}-t_{2})q_{2}^3p_{2}^2-2(t_{2}-t_{1})q_{1}q_{2}^2p_{1}p_{2}\\[10pt]
\phantom{H_{2}=}+(t_{1}-t_{2})q_{1}^2q_{2}p_{1}^2+(t_{2}+1)(t_{2}-t_{1})q_{2}^2p_{2}^2-2t_{2}(t_{1}-1)q_{1}q_{2}p_{1}p_{2}+t_{1}(t_{2}-1)q_{1}q_{2}p_{1}^2\\[10pt]
\phantom{H_{2}=}+t_{2}(t_{2}-1)q_{1}q_{2}p_{2}^2+(h-\alpha_{1}-\alpha_{2})(t_{2}-t_{1})(q_{1}q_{2}p_{1}+q_{2}^2p_{2})-t_{2}(t_{2}-t_{1})q_{2}p_{2}^2\\[10pt]
\phantom{H_{2}=}+\alpha_{4}t_{2}((t_{1}-1)q_{1}p_{1}-(t_{2}-1)q_{1}p_{2})-\alpha_{3}t_{1}(t_{2}-1)q_{2}p_{1}\\[10pt]
\phantom{H_{2}=}+(h(t_{1}-t_{2})+(\alpha_{1}+\alpha_{2})(t_{2}-t_{1})+\alpha_{3}t_{1}(t_{2}-t_{1})+\alpha_{4}t_{2}(t_{1}-t_{2})-\alpha_{5}(t_{1}+t_{2}^2-t_{2}-t_{1}t_{2}))q_{2}p_{2}\\[10pt]
\phantom{H_{2}=}-\alpha_{1}\alpha_{2}(t_{2}-t_{1})q_{2}+\alpha_{4}t_{2}(t_{2}-t_{1})p_{2}\}.
\label{hg11111t2}
\end{array}
\end{equation}   

\item
The case of \rm G(1,1,1,2)\\[4pt]
{\bf The Hamiltonian $H_1$ for $t_{1}$-flow.}
\begin{equation}
\begin{array}{l}
H_{1}=\dfrac{1}{(h+\alpha_{1}+\alpha_{2}+\alpha_{3}+\alpha_{4}+\alpha_{5})t_{1}^2}\{q_{1}^3p_{1}^2+2q_{1}^2q_{2}p_{1}p_{2}+q_{1}q_{2}^2p_{2}^2-t_{1}q_{1}^2p_{1}^2-t_{2}q_{1}q_{2}p_{2}^2\\[8pt]
\phantom{H_{1}=}+(\alpha_{3}+\alpha_{4}-h)(q_{1}^2p_{1}+q_{1}q_{2}p_{2})+(\eta +(2h+\alpha_{1})t_{1})q_{1}p_{1}+\alpha_{2}t_{2}q_{1}p_{2}+\eta t_{1}q_{2}p_{1}\\[8pt]
\phantom{H_{1}=}+\eta (1-t_{2})q_{2}p_{2}+\alpha_{3}\alpha_{4}q_{1}-\eta t_{1}p_{1}\}.
\label{hg1112t1}
\end{array}
\end{equation}  

{\bf The Hamiltonian $H_2$ for $t_{2}$-flow.}
\begin{equation}
\begin{array}{l}
H_{2}=\dfrac{1}{(h+\alpha_{1}+\alpha_{2}+\alpha_{3}+\alpha_{4}+\alpha_{5})t_{2}t_{1}(t_{2}-1)}\{t_{1}(q_{1}^2q_{2}p_{1}^2+q_{2}^3p_{2}^2+2q_{1}q_{2}^2p_{1}p_{2})\\[10pt]
\phantom{H_{2}=}+t_{2}(t_{2}-1)q_{1}q_{2}p_{2}^2-t_{1}(2t_{2}q_{1}q_{2}p_{1}p_{2}+(t_{2}+1)q_{2}^2p_{2}^2)+(\alpha_{3}+\alpha_{4}-h)t_{1}(q_{1}q_{2}p_{1}+q_{2}^2p_{2})\\[10pt]
\phantom{H_{2}=}+t_{1}t_{2}q_{2}p_{2}^2+\alpha_{2}t_{1}t_{2}q_{1}p_{1}+\alpha_{2}t_{2}(1-t_{2})q_{1}p_{2}+\eta t_{1}(1-t_{2})q_{2}p_{1}\\[10pt]
\phantom{H_{2}=}+(t_{1}(\alpha_{1}(t_{2}-1)+\alpha_{2}t_{2}-\alpha_{3}-\alpha_{4}+(2t_{2}-1)h)+\eta t_{2}(t_{2}-1))q_{2}p_{2}+\alpha_{3}\alpha_{4}t_{1}q_{2}-\alpha_{2}t_{1}t_{2}p_{2}\}.
\label{hg1112t2}
\end{array}
\end{equation}  

\item
The case of \rm G(1,1,3)\\[4pt]
{\bf The Hamiltonian $H_1$ for $t_{1}$-flow.}
\begin{equation}
\begin{array}{l}
H_{1}=\dfrac{1}{(h-\alpha_{1}-\alpha_{2}-\alpha_{3}-\alpha_{4})(t_{1}-t_{2})}\{q_{1}^2p_{1}p_{2}+q_{2}^2p_{1}p_{2}-2q_{1}q_{2}p_{1}p_{2}\\[10pt]
\phantom{H_{1}=}+(t_{1}-t_{2})(q_{1}^2p_{1}-2q_{1}p_{1}^2-2q_{2}p_{1}p_{2})+(2t_{1}^2-2t_{1}t_{2}-\alpha_{3})q_{1}p_{1}+\alpha_{2}q_{1}p_{2}\\[10pt]
\phantom{H_{1}=}+\alpha_{3}q_{2}p_{1}-\alpha_{2}q_{2}p_{2}+\alpha_{2}(t_{1}-t_{2})q_{1}+2\alpha_{1}(t_{1}-t_{2})p_{1}\}.
\label{hg113t1}
\end{array}
\end{equation}  

{\bf The Hamiltonian $H_2$ for $t_{2}$-flow.}
\begin{equation}
\begin{array}{l}
H_{2}=\dfrac{1}{(h-\alpha_{1}-\alpha_{2}-\alpha_{3}-\alpha_{4})(t_{2}-t_{1})}\{q_{1}^2p_{1}p_{2}+q_{2}^2p_{1}p_{2}-2q_{1}q_{2}p_{1}p_{2}\\[10pt]
\phantom{H_{2}=}-(t_{1}-t_{2})(q_{2}^2p_{2}-2q_{2}p_{2}^2-2q_{1}p_{1}p_{2})-\alpha_{3}q_{1}p_{1}+\alpha_{2}q_{1}p_{2}+\alpha_{3}q_{2}p_{1}\\[10pt]
\phantom{H_{2}=}-(2t_{1}t_{2}-2t_{2}^2+\alpha_{2})q_{2}p_{2}-\alpha_{3}(t_{1}-t_{2})q_{2}-2\alpha_{1}(t_{1}-t_{2})p_{2}\}.
\label{hg113t2}
\end{array}
\end{equation}  

\item
The case of \rm G(1,2,2)\\[4pt]
{\bf The Hamiltonian $H_1$ for $t_{1}$-flow.}
\begin{equation}
\begin{array}{l}
H_{1}=\dfrac{1}{(2h+\alpha_{1}+\alpha_{2}+\alpha_{3}+2\alpha_{4})t_{1}(t_{1}-t_{2})}\{(t_{1}-t_{2})q_{1}^2p_{1}^2+2t_{1}q_{1}q_{2}p_{1}p_{2}-t_{2}q_{1}^2p_{1}p_{2}-t_{1}q_{2}^2p_{1}p_{2}\\[10pt]
\phantom{H_{1}=}+(t_{2}-t_{1})q_{1}^2p_{1}+((\alpha_{1}+\alpha_{2}+\alpha_{3})t_{1}-(\alpha_{1}+\alpha_{2})t_{2})q_{1}p_{1}-\alpha_{1}t_{2}q_{1}p_{2}-\alpha_{3}t_{1}q_{2}p_{1}+\alpha_{1}tq_{2}p_{2})\\[8pt]
\phantom{H_{1}=}-\alpha_{1}(t_{1}-t_{2})q_{1}+t_{1}(t_{1}-t_{2})p_{1}\}.
\label{hg112t1}
\end{array}
\end{equation}  

{\bf The Hamiltonian $H_2$ for $t_{2}$-flow.}
\begin{equation}
\begin{array}{l}
H_{2}=\dfrac{1}{(2h+\alpha_{1}+\alpha_{2}+\alpha_{3}+2\alpha_{4})t_{1}(t_{1}-t_{2})}\{(t_{1}-t_{2})q_{2}^2p_{2}^2-2t_{2}q_{1}q_{2}p_{1}p_{2}+t_{2}q_{1}^2p_{1}p_{2}+t_{1}q_{1}^2p_{1}p_{2}\\[10pt]
\phantom{H_{2}=}+(t_{2}-t_{1})q_{2}^2p_{2}-\alpha_{3}t_{2}q_{1}p_{1}+\alpha_{1}t_{2}q_{1}p_{2}+\alpha_{3}t_{1}q_{2}p_{1}+((\alpha_{2}+\alpha_{3})t_{1}-(\alpha_{1}+\alpha_{2}+\alpha_{3})t_{2})q_{2}p_{2}\\[8pt]
\phantom{H_{2}=}-\alpha_{3}(t_{1}-t_{2})q_{2}+t_{2}(t_{1}-t_{2})p_{2}\}.
\label{hg112t2}
\end{array}
\end{equation}  

\item
The case of \rm G(1,4)\\[4pt]
{\bf The Hamiltonian $H_1$ for $t_{1}$-flow.}
\begin{equation}
\begin{array}{l}
H_{1}=\dfrac{1}{(2h+\alpha_{1}+\alpha_{2}+\alpha_{3})(t_{1}-t_{2})}\{-q_{1}^2p_{1}p_{2}-q_{2}^2p_{1}p_{2}+2q_{1}q_{2}p_{1}p_{2}+(t_{2}-t_{1})q_{1}^2p_{1}\\[10pt]
\phantom{H_{1}=}+\alpha_{2}q_{1}p_{1}-\alpha_{1}q_{1}p_{2}-\alpha_{2}q_{2}p_{1}+\alpha_{1}q_{2}p_{2}+\dfrac{1}{2}(t_{1}-t_{2})p_{1}(p_{1}-p_{2})\\[10pt]
\phantom{H_{1}=}-\alpha_{1}(t_{1}-t_{2})q_{1}-\dfrac{1}{2}t_{1}(t_{1}-t_{2})p_{1}\}.
\label{hg14t1}
\end{array}
\end{equation}  

{\bf The Hamiltonian $H_2$ for $t_{2}$-flow.}
\begin{equation}
\begin{array}{l}
H_{2}=\dfrac{1}{(2h+\alpha_{1}+\alpha_{2}+\alpha_{3})(t_{1}-t_{2})}\{q_{1}^2p_{1}p_{2}+q_{2}^2p_{1}p_{2}-2q_{1}q_{2}p_{1}p_{2}+(t_{2}-t_{1})q_{2}^2p_{2}\\[10pt]
\phantom{H_{1}=}-\alpha_{2}q_{1}p_{1}+\alpha_{1}q_{1}p_{2}+\alpha_{2}q_{2}p_{1}-\alpha_{1}q_{2}p_{2}+\dfrac{1}{2}(t_{1}-t_{2})(p_{1}p_{2}+p_{2}^2)\\[10pt]
\phantom{H_{1}=}-\alpha_{2}(t_{1}-t_{2})q_{2}-\dfrac{1}{2}t_{2}(t_{1}-t_{2})p_{2}\}.
\label{hg14t2}
\end{array}
\end{equation}  

\item
The case of \rm G(2,3)\\[4pt]
{\bf The Hamiltonian $H_1$ for $t_{1}$-flow.}
\begin{equation}
\begin{array}{l}
H_{1}=\dfrac{1}{(1+3h+2\alpha_{1}+2\alpha_{3})t_{1}}\{q_{2}^2p_{2}^2+\dfrac{1}{2}q_{1}q_{2}p_{1}+\dfrac{1}{2}q_{2}^2p_{2}-\eta t_{1}2q_{1}p_{2}\\[10pt]
\phantom{H_{1}=}+(1+h+2\alpha_{1}-\alpha_{2}+2\alpha_{3})q_{2}p_{2}+\dfrac{1}{2}\alpha_{1}q_{2}-\eta t_{1}(p_{1}-t_{2}p_{2})\}.
\label{hg23t1}
\end{array}
\end{equation}  

{\bf The Hamiltonian $H_2$ for $t_{2}$-flow.}
\begin{equation}
\begin{array}{l}
H_{2}=\dfrac{1}{(1+3h+2\alpha_{1}+2\alpha_{3})}\{-\dfrac{1}{2}q_{1}^2p_{1}-\dfrac{1}{2}q_{1}q_{2}p_{2}+2q_{2}p_{1}p_{2}+q_{1}p_{1}^2+\dfrac{1}{2}t_{2}q_{1}p_{1}+\dfrac{1}{2}q_{2}p_{1}-t_{2}p_{1}^2\\[10pt]
\phantom{H_{1}=}-\dfrac{1}{2}\alpha_{1}q_{1}+(1+h+2\alpha_{1}-\alpha_{2}+2\alpha_{3})p_{1}+\eta t_{1}p_{2}\}.
\label{hg23t2}
\end{array}
\end{equation}  

\item
The case of \rm G(5)\\[4pt]
{\bf The Hamiltonian $H_1$ for $t_{1}$-flow.}
\begin{equation}
\begin{array}{l}
H_{1}=\dfrac{1}{(3h+2\alpha_{1}-2\alpha_{2})}\{2q_{1}q_{2}p_{1}+q_{2}p_{1}^2+2q_{2}^2p_{2}-2q_{1}p_{2}+2t_{2}q_{2}p_{1}+2p_{1}p_{2}\\[10pt]
\phantom{H_{1}=}+2\alpha_{1}q_{2}+2t_{1}p_{1}+2t_{2}p_{2}\}.
\label{hg5t1}
\end{array}
\end{equation}  

{\bf The Hamiltonian $H_2$ for $t_{2}$-flow.}
\begin{equation}
\begin{array}{l}
H_{2}=\dfrac{1}{(3h+2\alpha_{1}-2\alpha_{2})}\{q_{2}^2p_{1}^2+2q_{1}^2p_{1}+2q_{1}q_{2}p_{2}-q_{1}p_{1}^2+2q_{2}p_{1}p_{2}+2t_{1}q_{2}p_{1}+2t_{2}q_{2}p_{2}\\[10pt]
\phantom{H_{1}=}-t_{2}p_{1}^2+p_{2}^2+2\alpha_{1}q_{1}-2t_{2}^2p_{1}+2t_{1}p_{2}\}.
\label{hg5t2}
\end{array}
\end{equation}  \end{enumerate}

These Hamiltonian systems with Hamiltonians \eqref{hg11111t1}-\eqref{hg5t2} obtained in this way may be called quantum Garnier systems in two variables.
For these systems, the following fact is important.

\begin{thm}
In each case, the obtained Hamiltonians $H_{i} (i=1, 2)$ $t_{1}$-flow and $t_{2}$-flow give commutative flow.
\end{thm}

\noindent {\it Proof.} 
The commutativity of two flows is equivalent to the following equation,
\begin{equation} \label{thm4.1-1}
[f, [H_1,H_2]-\frac{\partial H_1}{\partial t_{2}}-\frac{\partial H_2}{\partial t_{1}}]=0, \quad f=q_i, p_j.
\end{equation}
Indeed, we can show more stringent relations
\begin{equation} \label{thm4.1-2}
[H_1,H_2]=0, \quad \frac{\partial H_1}{\partial t_{2}}-\frac{\partial H_2}{\partial t_{1}}=0,
\end{equation}
by explicit calculation.
\qed

\medskip
In general, it is nontrivial to obtain quantum commutative expressions from classically commutative (Poisson commutative) ones.
In this paper, we succeeded it by imposing the condition of holomorphic properties. 
This result shows that the holomorphic property gives "good quantization".

\section{Summary and discussion}
In this paper, we constructed and characterized the quantum Garnier systems in two variables by holomorphic properties (\S3 Theorem \ref{thm}, \S4).
That is, for the Garnier systems G(1,1,1,1,1), G(1,1,1,2), G(1,1,3), G(1,2,2), G(1,4), G(2,3) and G(5), we succeeded their quantization by using the transformations constructed by Sasano and Suzuki.
For the quantization of Garnier systems, another approach has been studied \cite{N1,N2,NS} from the viewpoint of conformal field theory, where the KZ equation is considered to be a quantum Garnier system.
Comparison of this result with the present one is an interesting problem.

Finally, a possible direction of extensions of the result obtained is to extend it to multivariable cases.
Another direction is the extension to Sasano system whose  holomorphic properties were studied by Sasano \cite{S1}.

\section*{Acknowledgment}
I would like to express my sincere gratitude to Professor Yasuhiko Yamada for his valuable advices and encouragement during the course of this research.

\end{document}